\documentclass[a4paper]{article}
\usepackage{RR/RR}
\usepackage{hyperref}
\usepackage{epsfig}
\usepackage{cite}
\usepackage[tight,footnotesize]{subfigure}
\usepackage{url}
\usepackage{amsmath,amsthm}
\usepackage{color}

\usepackage{gensymb} 
\usepackage{algorithm2e}
\usepackage{xspace}

\newcommand{\INET}{{\itshape INET}}
\newcommand{\Omnet}{{\itshape OMNeT++}}

\RRdate{November 2009}

\RRauthor{
Paula Uribe\thanks{pauribejo@sophia.inria.fr}%
  \and
Juan-Carlos Maureira\thanks{jcmaurei@sophia.inria.fr}%
  \and
Olivier Dalle\thanks{olivier.dalle@sophia.inria.fr}%
}
\authorhead{Uribe, Maureira \& Dalle}

\RRtitle{
Extending INET Framework for Directional and Asymmetrical Wireless Communications
}
\RRetitle{
Extending INET Framework for Directional and Asymmetrical Wireless Communications
}
\titlehead{Directional and Asymmetrical Communications}
\RRresume{
  Ce rapport pr\'esente nos travaux pour \'etendre le cadre de simulations de 
  r\'eseau {\it INET} (bas\'e sur OMNET++) en lui ajoutant  un mod\`ele de
  radios directionnelles. Nous mettons ici l'accent particuli\`erement sur 
  l'impl\'ementation de communications asym\'etriques.  Dans un premier temps, nous 
  analysons la conception et les composants du mod\`ele actuel de radio d'{\it INET}. Dans 
  un second temps, nous en proposons des modifications qui permettent des communications 
  directionnelles. Nos r\'esultats pr\'eliminaires montrent que cette nouvelle extension 
  est assez flexible : elle permet \`a l'utilisateur, au prix d'un nombre limité de 
  calculs suppl\'ementaires, de modéliser n'importe quel forme de rayonnement de l'antenne.
}

\RRabstract{
	This paper reports our work on extending the \Omnet\, \INET\, 
	Framework with a directional radio model, putting a special
	emphasis on the implementation of asymmetrical communications. 
	We first analyze the original \INET\, radio model, focusing on its design and
	components. Then we discuss the modifications that have been done to support
	directional communications. Our preliminary results show that the new model is
	flexible enough to allow the user to provide any antenna pattern shape, 
	with only an additional reasonable computational cost.
}

\RRmotcle{
R\'esea sans fil, Handover, Simulation, Connectivit\'e sur la route, OMNeT++
}
\RRkeyword{
OMNeT++, INET Framework, Directional Radios, Asymmetrical communication
}
\RRprojets{Mascotte}
\RRtheme{\THCom} 
\RCSophia 

\begin{document}

\RRNo{7120}

\makeRR   

\section{Introduction}
\sloppy 

Mobile Ad-hoc Networks (MANETs) are receiving more attention in the last few
years. In many cases, the behavior of these networks depends on the performance 
of the radio communications in various directions around the emitters or receivers. 
This performance may be uniform in all directions, when using omni-directional antennas, 
or non-uniform, when using directional antennas. In the latter case, the antennas have a 
preferential direction of radiation, and a pattern of secondary emissions, at lower levels 
of radiation, in other directions. Thanks to such directional antennas, the network coverage 
can be increased and, at the same time, the use of less number of antennas
becomes possible, hence reducing the interference effects.\\
Since simulation tools are commonly used to design and evaluate the performance of such radio 
networks, simulation models of directional antennas are needed. Unfortunately, only a few 
simulation tools offer the ability to model directional antennas, such as OPNET\cite{opnet}, 
GloMoSim\cite{glomo}, NS-2\cite{ns2}, or QualNet\cite{qualnet}.
Indeed, modeling directional antennas is more complicated than modeling the omni-directional 
ones, because emitted or received power depends on the direction and a specific
attenuation pattern that depends on the antenna. These patterns often consist of a main
lobe in the direction of the antenna, and possibly some side and back lobes. The latter 
ones have a much lower amplification gain, but still cannot be neglected, in particular 
for the interferences calculation. In most of the simulators mentioned above, an
antenna's radiation pattern is described in a configuration file built by using mapping 
techniques. The gain values of the antenna are given for different points in space to 
determine the gain in all directions. However, there are some simulators that model the 
antenna using other techniques than mapping. In \cite{HarGoe2008}, Hardwick et al. use 
statistics to estimate the main and side lobes values for size and position, while 
in \cite{GhaBin2009}, Gharavi and Bin propose a 2D model, consisting of a ``pie-wedge'' 
pattern shape with circular constant gain for side and back lobes. In \cite{Kucuk07}, 
Kucuk et al. build a model of a smart antenna on top of \Omnet\,'s {\it Mobility 
Framework} for sensor networks applications, focusing on the performance 
evaluation, on the number of nodes, the traffic and the energy consumption.
\\
In this paper, we report on our work on extending the \Omnet\, \INET\,
Framework for supporting directional wireless communications. We propose
a {\it DirectionalRadio} module based on Gharavi and Bin's model\cite{GhaBin2009}, 
which allows the implementation of several antenna patterns. These patterns are 
defined by means of plugins that calculate the directional gain of the antenna 
in a two dimensional plane. Additionally, we present the modifications made 
to the \INET\, radio model in order to incorporate asymmetrical communications to 
support directional communications. We verify the correctness of the implementation 
by comparing the  antenna patterns observed in simulations against the expected 
theoretical patterns. We also evaluate our proposed radio model from the computational
point of view by providing quantitative data of the impact of including
asymmetrical communications in the overall execution time.

The paper is organized as follows: In Section \ref{sec:Status}, the current
\INET\, implementation and features are analyzed. Section
\ref{sec:ModelPresentation} depicts our proposed extension to the \INET\, radio
model for the directional antenna representation. Section \ref{sec:Implementation} 
presents the model implementation, its main features, parameters and limitations.
Section \ref{sec:Evaluation} provides a model evaluation by examining various
antenna patterns under two well defined scenarios. Finally, Section
\ref{sec:Conclusions} summarizes the proposed model, our main contributions and
further works towards the improvement of the \INET\,'s radio model.

\section{Status of INET Model}
\label{sec:Status}

\INET\, is an extensible network modeling framework that consists 
of a set of modules, built on top of the \Omnet\, simulator. It includes 
sub-models for several high-level protocols, such as Applications, IPv4, IPv6,
SCTP, TCP, UDP, and wired and wireless physical layers.
\begin{figure}[t]
\centering
\includegraphics[scale=0.29]{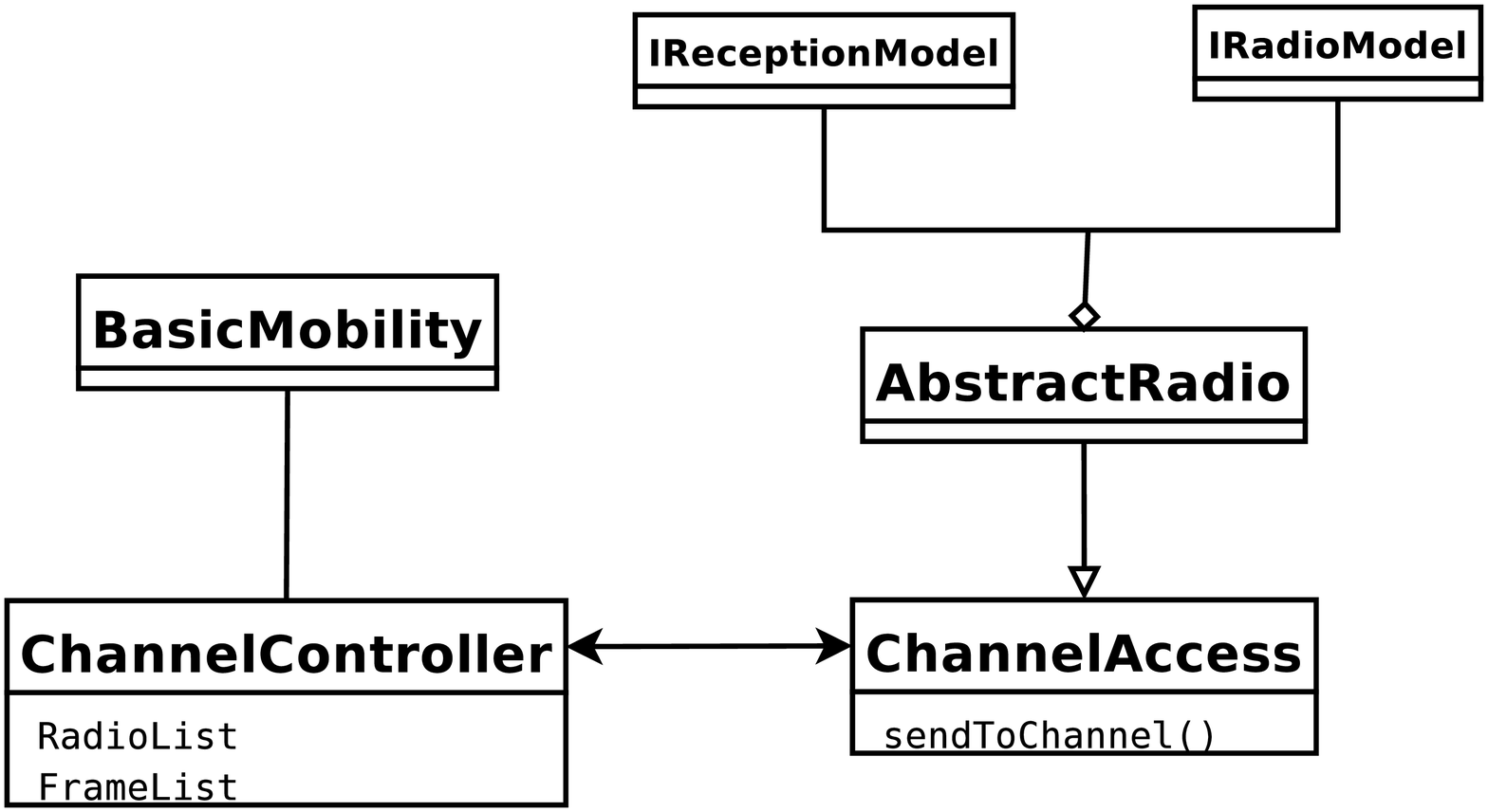}
\caption{INET's radio model components.}
\label{fig:modules}
\end{figure}
The initial design of the \INET\,'s radio model integrates an abstraction of
the channel, called the {\it ChannelController}, with a {\it ChannelAccess}
interface, which allows the {\it AbstractRadio} to interact with the
radio channel. Radios are built on-top of this {\it ChannelAccess}. Also, a {\it
Mobility} component is provided, giving nodes (with radio devices inside) the 
ability to move on the simulation playground. Fig. \ref{fig:modules} is a class
diagram of \INET\,'s radio model components.

The {\it ChannelController} is an abstraction of the radio channels. Its main 
roles are: being aware of all communications (packets in the air on each channel) 
that are happening; registering all the simulated physical radios and their positions;
and determining which radios are ``connected'', meaning which hosts are able to
communicate. This latter operation defines a connectivity graph, built by using
the Maximum Interference Distance to determine whether or not a radio is able to
``hear'' another radio, based on the Free Space propagation (Path Loss) defined 
by a set of identical radio parameters for all radios.

The {\it ChannelAccess} is an interface that provides the services
needed to send an airframe by the channel. This interface is the
cornerstone for implementing any radio. It is important to note that
{\it ChannelAccess} interacts with the {\it ChannelController}, both being
responsible for providing the required information to the radio in order to
decide if an airframe is correctly received or transmitted.

The {\it AbstractRadio} is an extension of the {\it ChannelAccess} module. It
provides a generic radio functionality, through a {\it RadioModel}
interface, which is responsible for deciding if a packet is correctly received, 
based on the noise measures. It also includes a {\it ReceptionModel} interface, 
in charge of calculating the airframe's reception power, using a propagation
loss model. There are two propagation models already implemented: the classical
{\it Free Space Pathloss} and the {\it Two-ray} with ground reflection.

Hosts (or nodes) contain one (or several) radio modules and one of the
available {\it Mobility} modules (circular, linear, random, etc.).

The {\it ChannelController} keeps the important information needed for the {\it
AbstractRadio}. Therefore, the {\it ChannelAccess} must provide means for 
the radio to access that information, since it is the entity that
connects both sides. Moreover, as mentioned above, the {\it ChannelController}
must build the connectivity graph (or hosts neighbors list),
that indicates with which nodes communications can occur. This graph varies 
only in response to a mobility event, which implies that the module that
triggers the graph's update is the {\it BasicMobility}, through the {\it
ChannelController}. In the current version of \INET\,, the communications follow 
a symmetrical model: For a pair of hosts $h1$ and $h2$, $h2$ is considered the 
neighbor of $h1$ if it is within its maximum interference distance, which, 
reciprocally, always implies that $h1$ is neighbor of $h2$ (the {\it ChannelController}
uses the same maximum transmission power for all radios, when calculating their maximum 
interference distance).

Currently\footnote{as of November 2009.}, there are several \INET\, branches
that introduce improvements in different ways. In particular, the branch supporting 
multiple radios assigns a neighbors list to each host's radio instead of to a
single list to the host. Therefore, each host has as many lists as radios. This change 
implies that the responsibility for building and updating the connectivity 
graph is transferred from the host to the radios of the host when a mobility event 
occurs.

\section{Extensions to INET's Radio Model}
\label{sec:ModelPresentation}

In this section, we introduce our proposed extension to \INET\,'s radio model to 
support directional wireless communications. We start with some technical concepts, 
followed by the description of our proposed extension. Then, we present an extended 
module for the \INET\, model, that supports directional and omni-directional communications.
Notice that the problem of asymmetrical communications addressed hereafter is deeper 
studied in Section \ref{sec:Implementation}.

\subsection{Concepts}

Antenna patterns are commonly used when studying propagation of radio signals
in the space. The {\it antenna pattern}\cite{Balanis} is a polar chart that 
describes the dependence of the radiation power (usually in relative decibels,
dB) and the direction of communication in 360\degree. For omni-directional antennas, 
the theoretical pattern is a circle. For directional antennas, it is an 
irregular shape, having in general a principal or main lobe and side/back
lobes. For directional patterns, the main lobe corresponds to the region 
where the largest amount of power is radiated, and can be characterized by its
maximum gain and its {\it Beam Width}, corresponding to the beam wideness, often defined 
by a 3dB-threshold\cite{Carr} that delimits it. Side/back lobes regions can
also be distinguished, where smaller amounts of power are radiated (generally seen as losses) 
with no preferential direction. These losses are often characterized by
their maximum gain. In order to quantify the signal propagation, the {\it Link 
Budget}\cite{rapp} equation is used. This equation corresponds to an account
of the transmission power, antennas gains and losses. The main contributions are 
the transmission power, antennas gains and free space propagation loss.

\subsection{Proposed Extension for the Radio Model}

Our proposed extension to the \INET\,'s radio model allows representation of
any antenna pattern, and specifically, to represent any gain function when 
calculating the {\it Link Budget} for a wireless communication. Our extension
separates the {\it Link Budget} into two phases: the calculation of the
antenna's gain in a pluggable external module, and the calculation of the effective reception 
power on the receiving end. In this model, the transmission and reception gain 
patterns are assumed to be identical, due to the Reciprocity theorem\cite{kraus}. 
This helps to apply the same procedure for the gain calculation, both for 
transmitting and receiving an airframe. In more detail, the external 
pluggable module is an abstraction of an antenna pattern, that gives the 
gain value in the direction of two-nodes communication. Our proposed extension 
uses this external module to calculate the {\it Link Budget} when determining 
the {\it effective transmission power} of an airframe. For two wireless 
hosts, the communication angle can be calculated from the transmitter's and 
receiver's coordinates, and consequently, the antenna gain in that direction can 
be determined. The {\it effective transmission power} is obtained by adding 
this gain and the nominal antenna transmission power. A similar calculation
is performed to determine the {\it effective reception power}, but using the
received power (i.e. the difference between the effective transmission power
and the path loss) instead of the nominal transmission power. More formally,
for omni-directional communications, the simplified {\it Link Budget}
expression is as follows:
\begin{eqnarray}
P_{rx} &=& P_{tx} - PL 
\label{eq:linkbudget-1}
\end{eqnarray}
when considering the antenna gains for directional comunication, the expresion
becomes as:
\begin{eqnarray}
 P_{rx} &=& P_{tx} + G_{tx} - PL + G_{rx}
\label{eq:linkbudget-2}
\end{eqnarray}
where $P_{tx}$ is the nominal transmission power, $G_{tx}$ is the transmitter
gain, $G_{rx}$ is the receiver gain, $PL$ is the path loss and $P_{rx}$ is the
effective received power.

The advantages of separating the antenna gain from the {\it Link Budget} is the 
flexibility introduced by the externalization of the antenna gain calculation, 
allowing us to implement any antenna pattern without changing the radio functionalities. 
Directional and omni-directional antenna patterns can be implemented by describing 
the antenna gain with mathematical curves, as well as by mapping techniques. We used 
a simplified {\it Link Budget} expression that only considers the antenna gains and 
propagation losses, as expressed in Equation \ref{eq:linkbudget-2}, while
losses due to cables, connectors, or any other type of losses are not considered.

\begin{figure}[!th]
\centering
\includegraphics[scale=0.20]{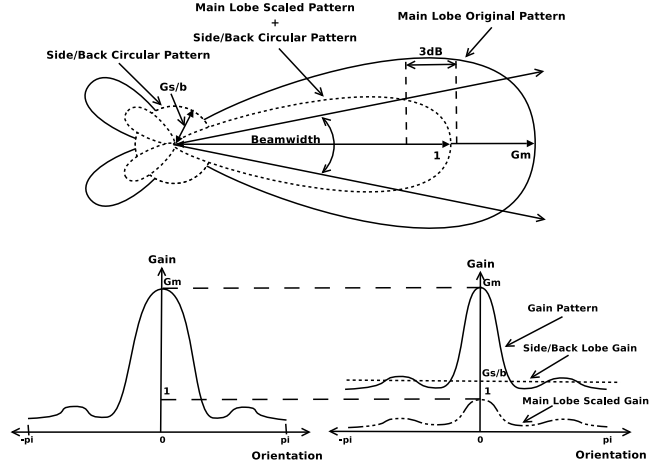}
\caption{Example of antenna pattern and the scaled antenna pattern.}
\label{fig:scaled2}
\end{figure}

We use the previous described extensions to implement four directional antenna
patterns, based on the ``pie-wedge'' model described by Gharavi et al. in
\cite{GhaBin2009}. The radiation pattern is represented by two main components: a main lobe and
side/back lobes. We used different mathematical curves (folium, cardioid, circle and rose) to
analytically define the main lobe's gain function in every direction (from 0\degree to 
360\degree); and a circle with an unity-gain (onmi-directional communication) to 
represent the side/back lobes. Then, while the gain of the
main lobe varies according to the mathematical formula evaluated in
different points, defined by the transmission/reception angle, the
gain of the side/back lobes remains constant. For practical reasons, we
scale the original analytical curve that defines the main lobe gain pattern, in
order to obtain a curve with a maximum gain of $G_{m}$ in the main beam direction. 
Also, the side/back lobes are represented by a circular unity-gain
pattern, which multiplied by the side/back lobes gain, gives the circular pattern 
with radius $G_{s/b}$. Fig. \ref{fig:scaled2} depicts the original and the scaled 
analytical curves used by the proposed directional radio, where we observe the gain 
function before and after the scaling. Note that the maximum gain remains the same, 
but now the gain function corresponds to the maximum between the main lobe scaled gain 
and the side/back lobes gain. This means that within the area defined by the 3dB-threshold, 
the main lobe gain dominates, while within the rest of the space, the gain value
alternates between the main lobe scaled curve and the the side/back lobes circle.

\section{Model Implementation}
\label{sec:Implementation}

In this Section, we describe the implementation of our proposed directional
radio module. This implementation can be separated into two different parts:
i) the modifications that have to be made to current \INET\, version when 
implementing asymmetrical communications; and, ii) the changes made to 
incorporate directional communications, including four different
antenna patterns and the proposed {\it Link Budget} calculation presented in the
previous section.
\\
\\
\subsection{Asymmetrical Communications}
\label{subsec:AsymmComm}

Our \INET\, branch extends the multiple radios branch to support
asymmetrical communications, where radios no longer are assumed to have
the same antenna pattern and transmission power. It also allows the user to select one
of the implemented antenna patterns, or create a new one. This
improvement forces to change the way the neighbor lists are calculated 
and updated, since the premise {\it if you hear me, I can hear you}, is 
not always true\cite{Kotz04} when considering asymmetrical communications. 

The first step to implement asymmetrical communications on the current
\INET\, radio model, is to relieve the {\it ChannelController} of the
responsibility of calculating the maximum interference distance. According to
an asymmetrical radio model, radios may have different coverages (in range and 
in shape). Thus, this task should be assigned to the radio itself,
implementing it at the {\it AbstractRadio} level. Consequently, the {\it
ChannelAccess} interface would rely on the {\it ChannelController} to build 
the neighbors list, but now using a method provided by the radio to 
determine when a node is under its radio coverage. This
method, called {\it isInCoverageArea}, relies on the {\it ReceptionModel} to 
determine the maximum interference distance, according to the antenna's radiation 
shape, and also to the radio signal propagation model.

Regarding the connectivity graph determination, the event that triggers an update 
in our implementation is not only a mobility event, but a mobility event or a 
transmission request, since the neighbor list of any node {\bf must} be updated
when transmitting a packet, and so, be faithful with the radio signals that
each node receives. Additionally, this update is now related to all
nodes (and radios) in the simulation, and not only to the node that has changed its position. 
These modifications have a detrimental impact on the overall execution time.
Firstly, the neighbor discovery algorithm has now a complexity of $O(n^2 m)$,
$n$ being the number of hosts, and $m$ the number of radios per host; and
secondly, this algorithm is executed more often. Preliminary results shown that
the computational cost is higher enough to make the model not attractive to 
users when using this algorithm of ``brute force'' to update the neighbor lists. 
So, we developed an alternative algorithm that exploits the locality of nodes 
to calculate their neighbors lists and supports different coverage ranges. 
In Section \ref{sec:performance}, we analyze the computational cost of both 
algorithms in terms of the execution time. The next section presents the 
proposed algorithm for the neighbor list calculation.

\subsection{NeighborsGraph Algorithm}
\label{subsub:NeighborsAlg}

In the previous section, we showed the need for an efficient algorithm
to overcome the increase of the model's execution time when using asymmetrical
communications. This algorithm should exploit the locality of nodes, since
when a single node moves, let's say node $N_{1}$, the neighbors lists that must
be updated are only those that change due to $N_{1}$'s movement, meaning 
the inclusion or the exclusion of $N_{1}$ as a neighbor. Updating a node that
is far away from $N_{1}$ is useless, in terms of the propagation model and
noise levels. This leads to our \emph{selective neighbor list
update algorithm}.

\begin{figure}[!ht]
\centering
\includegraphics[scale=0.27]{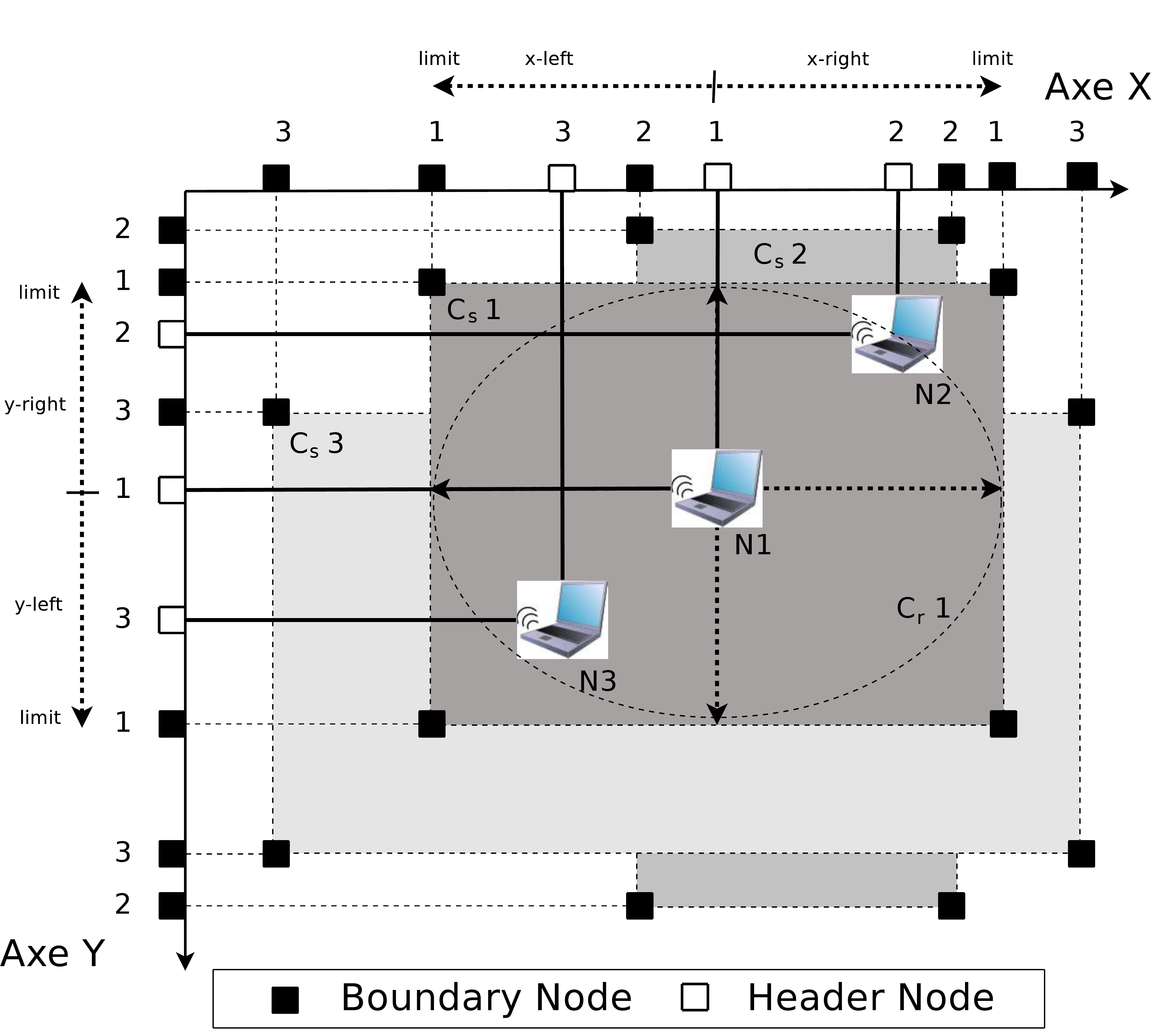}
\caption{NeighborsGraph data structure}
\label{fig:neighborgraph}
\end{figure}

This algorithm uses a data structure, similar to a sparse matrix, but keeping 
the positions of nodes (in fact, radio positions) and a coverage area (Figure 
\ref{fig:neighborgraph}). This area, $C_{s}$, is the square box that contains 
the real coverage area $C_{r}$. The boxed coverage area $C_{s}$ is defined by 
the maximum interference distance given a propagation model, when the received 
power is equal to the noise level. In the case of the Pathloss model, this 
distance depends on the transmitter power, carrier frequency and Pathloss coefficient. 
This distance defines $C_{s}$, which is used by the algorithm to compute the neighbors 
list within this region. Afterwards, a refinement of this list can be made by using the
$isInCoverageArea$ method of each radio to get the real neighbors list (nodes
within $C_{r}$). It is worth to notice that the $isInCoverageArea$ method is now
applied to a reduced set of nodes, instead of all nodes in the simulation.
Additionally, the neighbor list calculation returns the list of nodes to be
updated as a consequence of $N_{1}$'s displacement. For all those nodes, 
their neighbor lists are tagged as invalid, forcing them to be updated when
the node wants to transmit a packet.
The algorithm determines the neighbor list for a node $N_{1}$ by exploring the 
sparse matrix axes on each direction: {\bf x-left, y-right, x-right} and
{\bf y-left}; finding the {\it header nodes} within the region $C_{s}$. It also
determines the updated list by finding the {\it boundary nodes} that are traversed by
$N_{1}$. Each node's position (header and boundary nodes) is updated every time
the node moves in a $O(\log(n))$ operation, since the matrix axes are red-black
trees\cite{GuSe78}. A formal pseudo-code for the proposed algorithm is shown
in Algorithm \ref{alg:nl}. 

\incmargin{1.2em}
\linesnumbered
\dontprintsemicolon
\begin{algorithm}[ht]
\SetLine

\SetKwData{ds}{directions}
\SetKwData{axe}{axe}
\SetKwData{current}{current}
\SetKwData{direction}{direction}
\SetKwData{limit}{limit}
\SetKwData{n}{n}
\SetKwData{p}{p}
\SetKwData{nc}{n`}
\SetKwData{neighbors}{neighbors}
\SetKwData{toUpdate}{toUpdate}
\SetKwFunction{pos}{pos}
\SetKwFunction{prevpos}{prevpos}
\SetKwFunction{boundary}{boundary}
\SetKwFunction{owner}{owner}
\SetKwFunction{type}{type}
\SetKwFunction{inRange}{inRange}
\KwIn{node \n, empty \neighbors and \toUpdate lists}
\KwOut{updated \neighbors list and \toUpdate list}
\BlankLine
\Begin{
	\ds $\leftarrow$ \{ x-left,y-right,x-right,y-left \}\;
	\ForEach{\direction in \ds} {
		\axe $\leftarrow$ axe of \direction\; 
		\limit $\leftarrow$ \pos{\n,\axe} + \boundary{\n,\direction}\;
		\current $\leftarrow$ header node of \n on \axe\; 
		\While{ \pos{\current,\axe} not reach \limit} {
			\nc $\leftarrow$  \owner{\current}\;			
			\If{\type{\current} = header node }{
				\p $\leftarrow$ \pos{\nc,in the contrary \axe}\;
				\If{\inRange{\n, \p } }{
					add \nc to \neighbors \;
				}
			}
			\If{\type{\current} = boundary node}{
				b1 $\leftarrow$ \inRange{\nc,\pos{\n,\axe}}\;
				b2 $\leftarrow$ \inRange{\nc,\prevpos{\n,\axe}}\;
				\If{b1 $\neq$ b2}{
					add \nc to \toUpdate \;
				}
			}
			\current $\leftarrow$ next node in \direction\;
		} 
	}
	remove from \neighbors repeated nodes\;
	remove from \toUpdate repeated nodes\;
}

\caption{The NeighborsGraph Algorithm.}
\label{alg:nl}
\end{algorithm}
\decmargin{1.2em}

The function $pos(node,axe)$ returns the position of the node on the given axe;
$boundary(node,direction)$ returns the $C_{s}$ boundary limit in the given
direction; and $owner(node)$ returns the reference to the host/radio of the
node. As Figure \ref{fig:neighborgraph} shows, each host has a header
node, referencing the host position on each axe, and four boundary nodes,
referencing the limits of the region $C_{s}$. All these nodes have a direct
reference to their host/radio module. Finally, the function
$inRange(node,position)$ returns whether the given position is within
the $C_{s}$ region, or not. As we already mentioned, after the execution of this
algorithm, the real coverage of each radio inside the node $n$ is evaluated for
each neighbor host, in order to determine the real neighbors list. All neighbor
lists belonging to a host contained in the $toUpdate$ list are invalidated,
thus, forcing them to be updated when the host sends a packet to the channel.

The complexity of this algorithm may be higher than the brute force update
algorithm on a single host, since it visits each node at least two times (x and y 
directions). But, we improve the performance by doing this operation only for
the nodes affected when a certain node moves, instead of updating the neighbor
list of all nodes in the simulation. Performance evaluation results are given in 
Section \ref{sec:performance}.

\subsection{DirectionalRadio Module}
\label{subsec:directionalRM-Eval}

Based on the proposed extension of \INET\,'s radio model, we created the
{\it DirectionalRadio} module, which complements the current radio implementation 
with new functionality. In order to make this module flexible enough to allow
several antenna patterns, an {\it AntennaPattern} interface has been added. This 
interface contains the specific operations of a directional antenna and enables 
the implementation of new antenna patterns, without interfering with existing 
basic radio operations. 

As stated in Section \ref{sec:ModelPresentation}, the main lobe curve is scaled
to $1$ (i.e. its maximum value is $G_{m}$), according to the main lobe width as specified in the
configuration file. For a given direction of communication, the gain is given by
the higher value between the main lobe's and the side/back lobes' gain. 

We provide four antenna patterns, which use known mathematical curves: {\it
CircularPattern}, {\it CardioidPattern}, {\it FoliumPattern} and {\it
RosePattern}. These antenna patterns are customizable by changing the settings
in the configuration file.

There are some parameters common to all patterns, proper to directional
antennas: the {\it beamWidth}, the angular distance that indicates the
main lobe width, in degrees; the {\it mainLobeGain}, the maximum gain
of the main lobe, measured in dB; the {\it sideLobeGain}, the maximum
gain of the side/back lobes, in dBi; the {\it mainLobeOrientation}, which indicates
the direction at which the main lobe is pointing, in degrees; the {\it
dBThreshold}, the threshold value that defines the main lobe area, in
dB; and the {\it patternType}, the selected antenna pattern shape,
specified by its name. There are also some specific parameters to each case.
For example, if using the {\it CircularPattern}, the radio $r$ must be set, or
the $a$ and $b$ parameters for the {\it FoliumPattern}.

\section{Model Evaluation}
\label{sec:Evaluation}

We evaluated two aspects of our proposed directional radio module implementation: its 
correctness and its computational cost. The correctness of the model implementation is 
evaluated by comparing our simulation results with similar ones found in the literature. 
The computational cost is estimated by measuring the execution time of the same
simulation model with three different scenarios: a symmetrical model scenario, an asymmetrical model 
scenario with full update of neighbors, and an asymmetrical model scenario with the neighbors-graph 
algorithm to compute and update the neighbors.

\subsection{Correctness of the Proposed DirectionalRadio Module}

We designed two simulations to validate and evaluate the correctness of our implementation
of the {\it DirectionalRadio} module. The first one is intended to obtain and analyze 
the simulated antenna pattern. The second is intended to compare the performance of 
omni-directional and directional communications.

\subsubsection{Obtaining an Antenna Pattern}

This simulation has two objectives: to exhibit the antenna pattern, and to verify how 
it changes according to the transmitter-receiver distance and the path loss effect. 
The pattern obtained through the simulation is expected to match the settings given 
in the configuration file. The simulated scenario consists of one directional AP with
the main lobe oriented at $90\degree$, and $10$ omni-directional hosts following a circular 
path with different radiuses surrounding the AP. The simulation begins with all hosts 
aligned at $0\degree$ and separated by $10$ meters each, and ends when all hosts are 
back to their initial position after a complete revolution around the AP. For each host, 
each {\it Beacon's Received Power} is logged to assess whether the beacon was correctly 
received, or not.  

The {\it FoliumPattern} antenna was selected for this simulation. The 
radio parameters set for this simulation are shown below as an excerpt 
from the configuration file.

\begin{figure}[th!]
\centering
\includegraphics[scale=0.2]{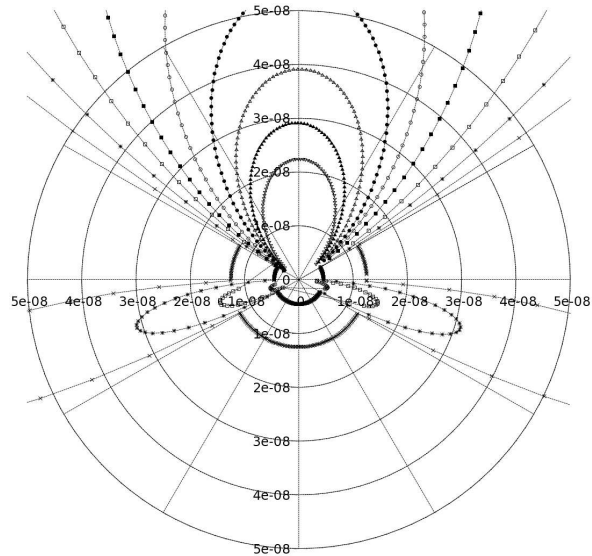}
\caption{Observed folium pattern at different distances transmitter/receiver
obtained from the simulation.}
\label{fig:levelcurves}
\end{figure}

\scriptsize
\begin{verbatim}
# Antenna Pattern Parameters
**.ap1.wlan.radio.transmitterPower = 40.0mW
**.ap1.wlan.radio.beamWidth = 40deg
**.ap1.wlan.radio.mainLobeGain = 15dB
**.ap1.wlan.radio.sideLobeGain = -5dBi
**.ap1.wlan.radio.mainLobeOrientation = 90deg
**.ap1.wlan.radio.dBThreshold = 3dB

# Folium Pattern
**.ap1.wlan.radio.patternType = "FoliumPattern"
**.ap1.wlan.radio.FoliumPattern.a = 1
**.ap1.wlan.radio.FoliumPattern.b = 3
\end{verbatim}
\normalsize
Fig. \ref{fig:levelcurves} shows the observed antenna pattern for different
distances between transmitter and receiver. The angular axis corresponds to 
the angle between the AP and the hosts, while the radial axis corresponds to
the received power in milliwatts (mW). 

We successfully verified that the antenna pattern matches with the set folium curve and its
orientation. Regarding the side/back lobes, we can observe how the gain value
alternates between the folium curve and the circle, depending on what is the
higher value, as described in the implementation section. We can observe that
the received power varies according to the distance between transmitter and
receiver due to the path loss effect. Following one radial direction, the difference
between two consecutive level curves corresponds to the path loss when the
radio signal travels $10$ meters.

\subsubsection{Omni-directional vs. Directional Communications}

\begin{figure}[!th]
\centering
\includegraphics[scale=0.2]{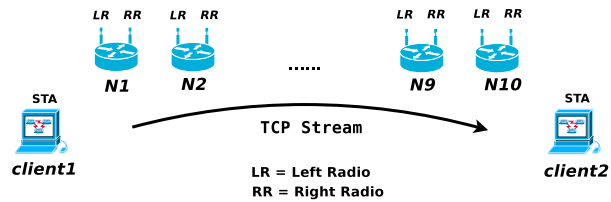}
\caption{Second evaluation scenario. Mesh network with multiple radio
interfaces nodes.}
\label{fig:meshschema}
\end{figure}

The goal of the second simulation is to evaluate our {\it DirectionalRadio} 
by comparing the network throughput when using omni-directional and
directional antennas. For this purpose, we build a mesh network simulation,
with $10$ nodes containing $2$ bridged radio interfaces, and $2$ client hosts
with a single radio interface. In this scenario, the radio parameters for the hosts 
are chosen such that directional antennas are able to ``hear'' up to $5$ antennas placed
in their same direction of orientation, even when they are supposed to communicate
only with their immediate neighbors. We determined that using a network with
$10$ hosts is enough to create strong interference between antennas and to clearly
observe the behavior of communications. All antennas in the simulation have
the same maximum transmission range, and the radio interfaces are configured as
shown in Fig. \ref{fig:meshschema}, using the same channel. In this scenario, a 
TCP stream is transmitted first time from $client1$ to $client2$ through the mesh network,
using omni-directional antennas (1st case) and then, a second time using
directional antennas (2nd case); network throughput was monitored during the experiment.

\begin{figure}[!th]
\centering
\includegraphics[scale=0.65]{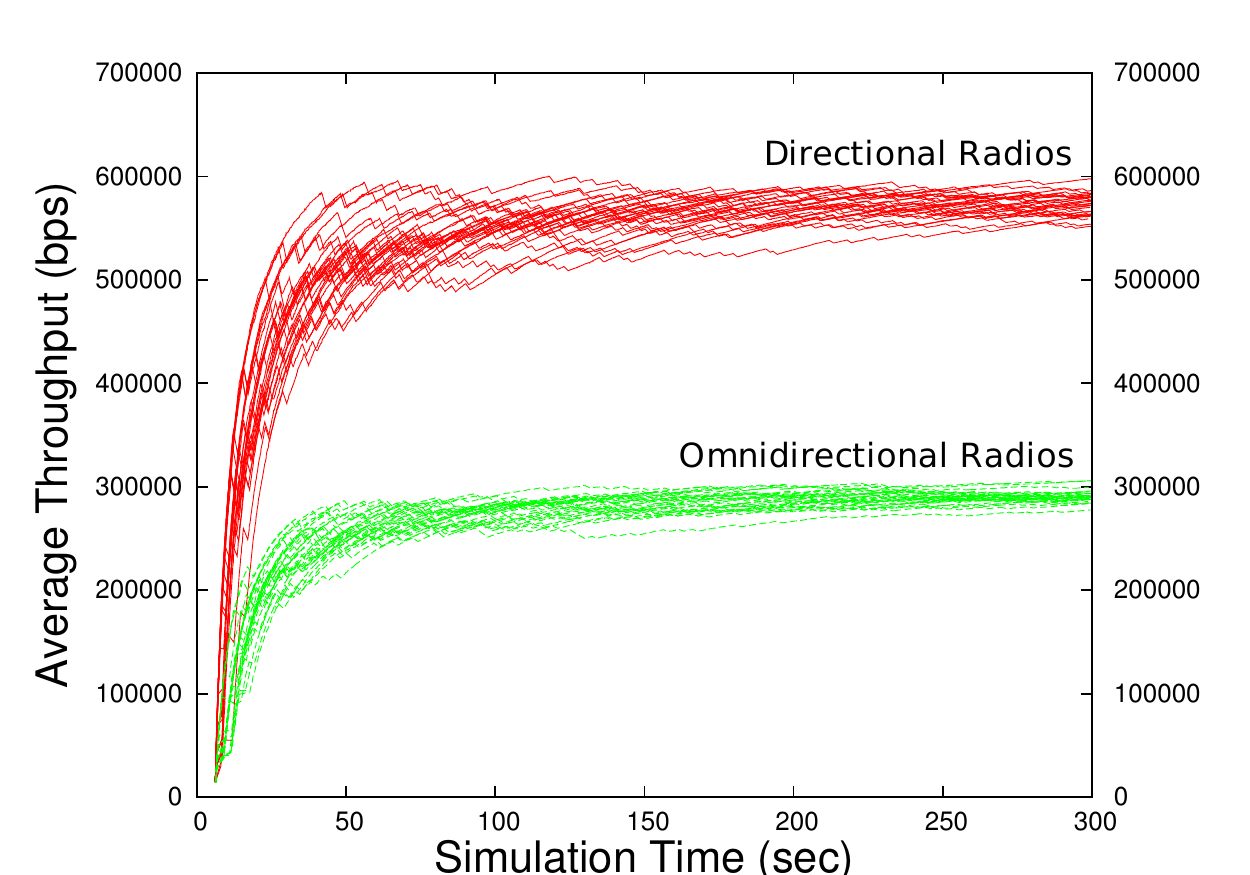}
\caption{Comparison of TCP throughput for the 1st and 2nd case for 30
repetitions of the simulation.}
\label{fig:mesh}
\end{figure}

\begin{figure}[!ht]
\centering
\subfigure[Omnidirectional radios]{
	\includegraphics[scale=0.55]{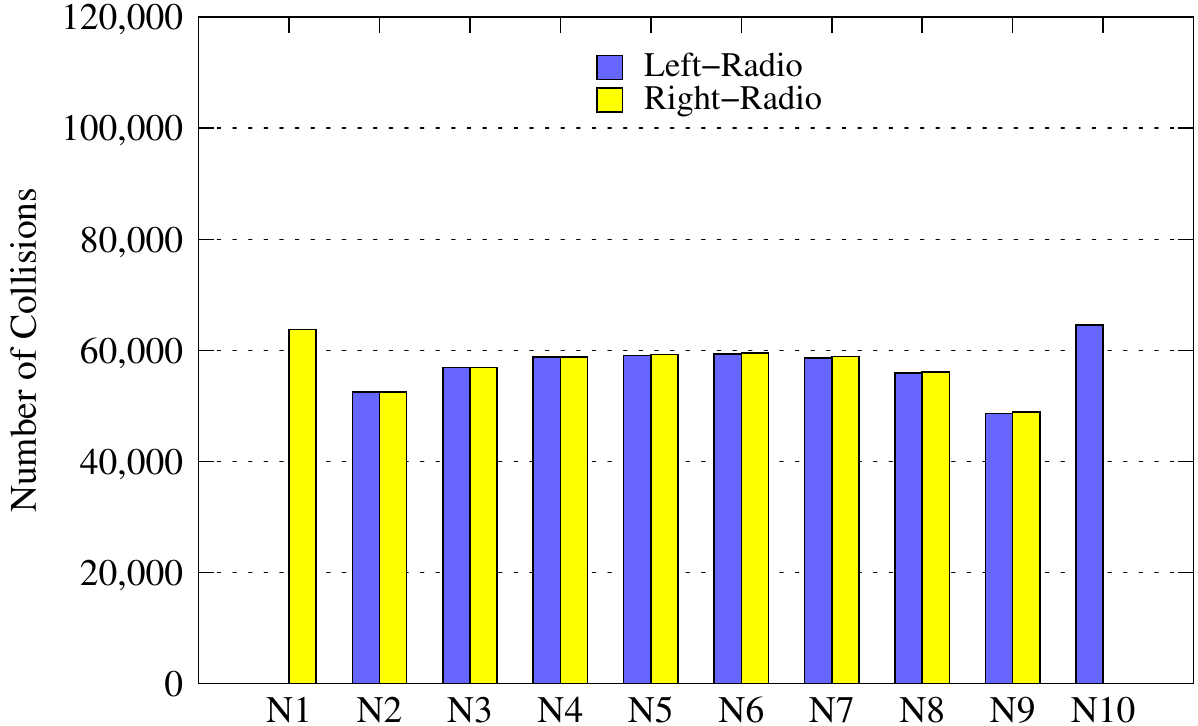}
	\label{collisions-a}
}
\subfigure[Directional radios]{
	\includegraphics[scale=0.55]{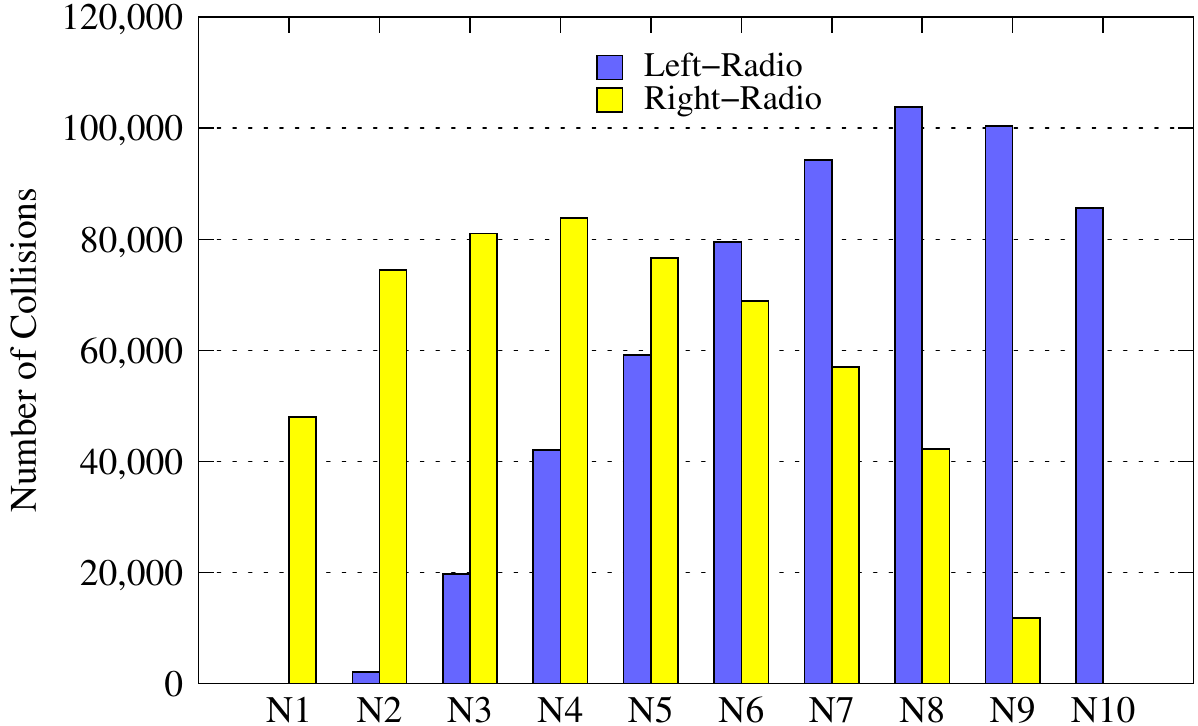}
	\label{collisions-b}
}
\caption{Comparison for packet collisions for the 1st and 2nd case.}
\end{figure}

\begin{figure}[!ht]
\centering
\subfigure[Omnidirectional radios]{
	\includegraphics[scale=0.55]{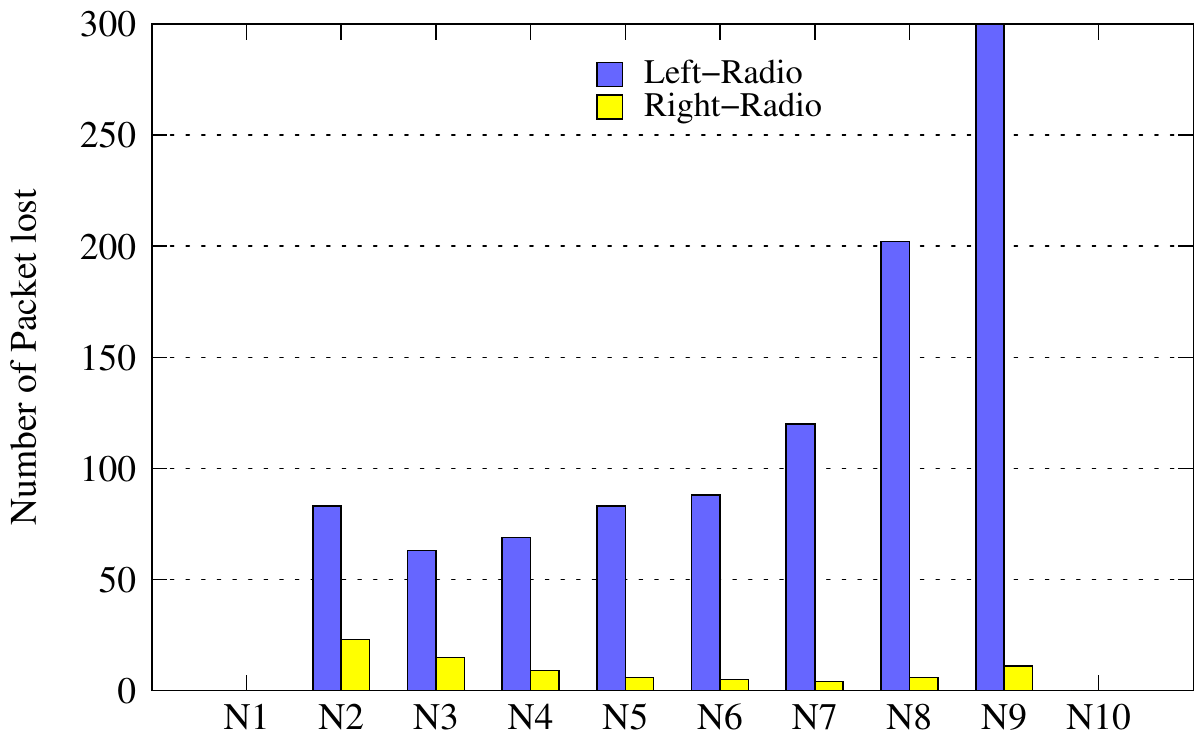}
	\label{pkt-loss-a}
}
\subfigure[Directional radios]{
	\includegraphics[scale=0.55]{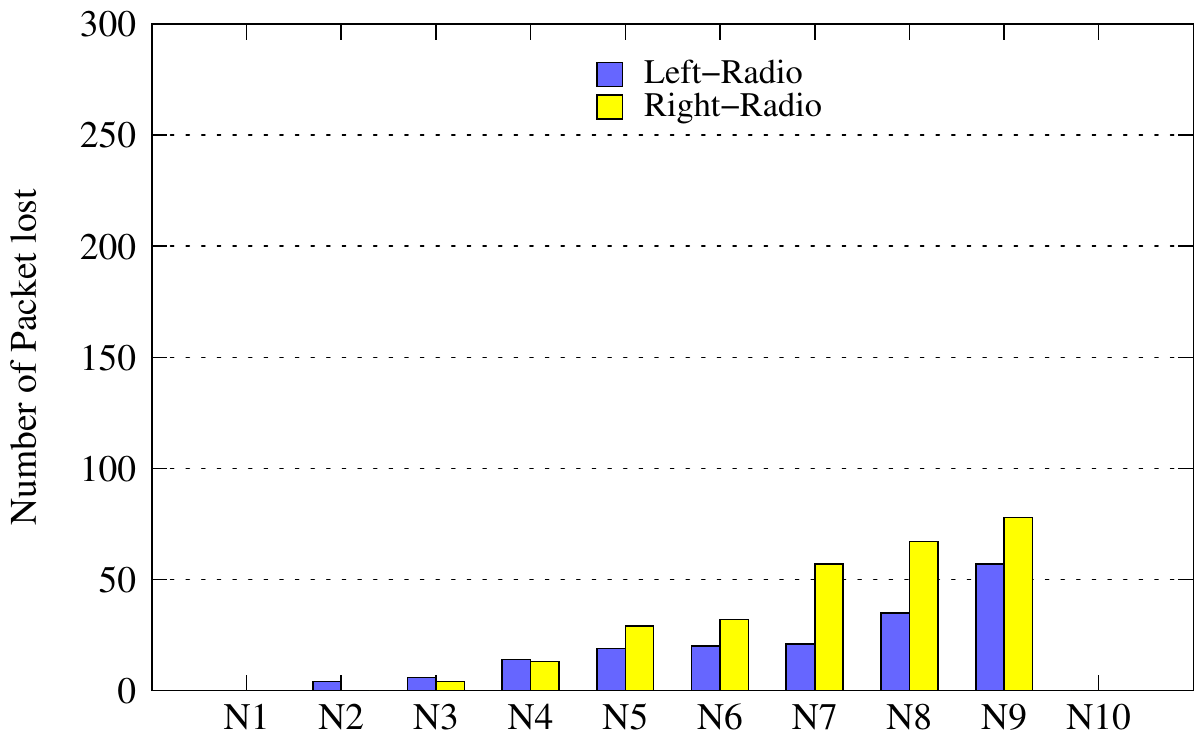}
	\label{pkt-loss-b}
}
\caption{Comparison for packet losses for the 1st and 2nd case.}
\end{figure}

Fig. \ref{fig:mesh} presents a comparison of the network throughput for the two
cases described above. In the 1st case, we observe that the network
throughput is about half of the theoretical one, which is consistent with
previously obtained results \cite{throughput}. In Fig. \ref{collisions-a} we can
see that the number of collisions using omni-directional nodes is constant for
all radios, while for directional antennas (Fig. \ref{collisions-b}) the number
varies depending on the node's position. This is because the coverage range
for directional antennas is set to be greater, thus allowing two distant
antennas to ``hear'' each other. The left radio of $N8$, for instance, is able
to hear from $N2$ to $N7$, leading to a greater number of collisions when
receiving. The packet loss, however, is reduced when using directional antennas,
as shown in Fig. \ref{pkt-loss-b}, compared to the case that uses
omni-directional antennas (Fig. \ref{pkt-loss-a}).

\subsection{Computational cost Analysis}
\label{sec:performance}

The introduction of asymmetrical communications generates an increase in the
execution time of the neighbors discovery procedure, since it can not be
assumed reciprocity when building (or updating) the neighbor list. However,
the computational cost can be partially decreased if a more refined algorithm is
used. In order to evaluate the performance of each algorithm, we compare
the simulation time for the same scenario when using: the current
neighbor discovery procedure described in Section \ref{sec:Status} for a
symmetrical model (Procedure 1); the ``brute force'' neighbors discovery
procedure for the asymmetrical model described in Section \ref{subsec:AsymmComm} 
(Procedure 2); and the {\it NeighborsGraph Algorithm} described in Section
\ref{subsub:NeighborsAlg} (Procedure 3).

The simulation scenario consists of 100 hosts, randomly moving at different
speeds (up to $40$ Km/h). There are $4$ access points belonging to the same
network, located in a 2$\times$2 grid, covering all the simulation playground and
connected to a server via wired links. All radios in simulation (APs and
hosts) have the same radio parameters. The hosts are expected to associate to the
network performing handover when passing from one AP's coverage area to
another's. All along the simulation, hosts are sending ICMP traffic (ping) to
the server. The simulation time was set to 500 seconds and 10 repetitions for
each model were ran, logging the real execution time for each case. The hardware
used to simulate this scenario was a Dell Precision T3400 (Intel(R) Core(TM)2
Duo CPU E6550 @ 2.33GHz, 2Gb RAM running Fedora 11).

\begin{figure}[th!]
\centering
\includegraphics[scale=0.44]{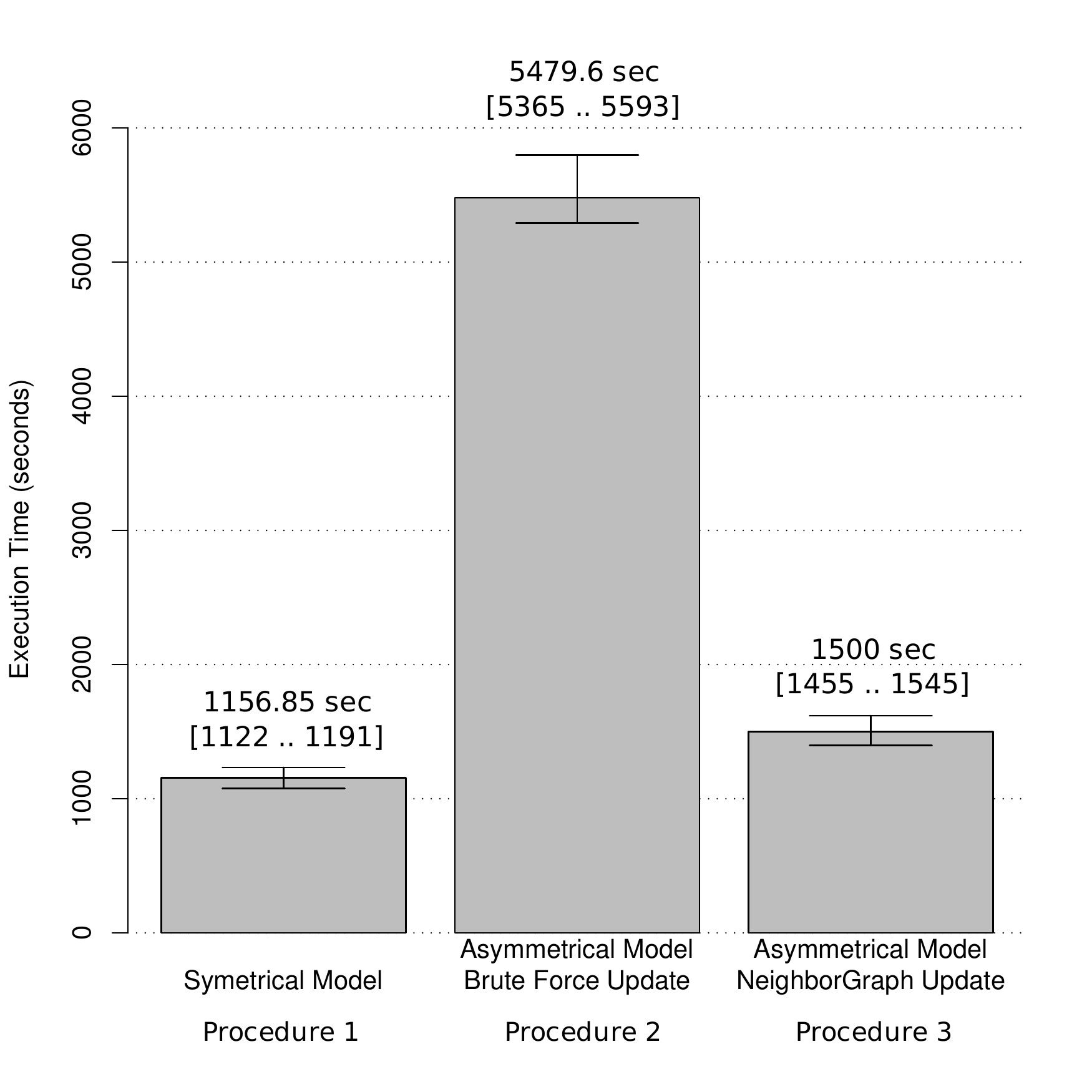}
\caption{Execution time for Symmetrical and Asymmetrical model with different
neighbors list update algorithms.}
\label{fig:exectime}
\end{figure}

Figure \ref{fig:exectime} shows the execution time for the Procedures 1, 2 and
3, respectively. We can observe that using Procedure 2 increases $\approx$
500\% the computation time compared with Procedure 1, while the increase is close 
to a 50\% when using Procedure 3. A comparison of means\cite{Montgomery2006}
among the procedures shows that, while the difference between Procedure 1 and
Procedure 2 is always large, the difference between Procedure 1 and
Procedure 3 always remains about 50\%.  

We realize that Procedure 1 takes advantage from the assumption of symmetry of communications when
determining the nodes to be updated, having a very good performance in execution time.
Contrarily, when asymmetrical communications are used, the neighbors list must 
be updated much more often to honor the SNR and noise calculation and this is done for 
all nodes when a host moves or transmits an airframe. Thus, the execution time 
is increased dramatically. Nevertheless, our proposed algorithm to calculate the
neighbors overcomes this impact, reducing the increase in execution time reasonably 
when using an asymmetrical communications radio model.

Extending these results to larger simulations, we estimate that the execution
time of our proposed algorithm will be always higher than the execution time
when symmetrical communications are used. However, further experimentation
has shown that this difference is always around 50\%. As we stated in
Section \ref{subsub:NeighborsAlg}, the execution time of the neighbors
graph algorithm is expected to be higher since each neighbor node is visited
at least twice (axe-x and axe-y). But, a possible alternative to overcome this
overhead could be to perform this exploration in a parallel way (exploiting the
multi-core architectures). Whether or not the execution time of the neighbors graph
algorithm could be improved, it will never be better than the algorithm for
symmetrical communications, since the assumption of symmetry gives the best
case when all radio coverages are equal. On the contrary, our algorithm 
ensures a good approximation of the optimal case when calculating the neighbors
lists for a node, not only when radios coverages are equal, but also in cases where
radios coverages are different in shape and size.

\section{Conclusions}
\label{sec:Conclusions}

In this work, we presented an extension of the \Omnet\,'s
\INET\, Framework for directional and asymmetrical wireless communications.
This new extension is based on the existing \INET\, multi-radio branch and introduces
the following contributions: an asymmetrical Radio Model to support directional 
antenna patterns; a {\it NeighborsGraph} Algorithm to speed up the neighbor
nodes update computation; and an implementation of a Directional Radio module
with four antenna patterns. The included antenna patterns use mathematical
curves to represent the gain pattern in a pie-wedge directional radiation model.

Although the proposed model still has several simplifications, we demonstrated through 
simulations that the results obtained agree with the results found in the literature 
when using directional radios in mesh networks.  

Furthermore, despite the increase in execution time could be potentially high when
using asymmetrical communications, we showed it is possible to
reasonably reduce it by using our {\it NeighborsGraph} algorithm.

In this work, we also identified open issues, such as the accuracy of antenna gain
in the plane, or the use of multi-core architectures to
speed-up the construction/update of the neighbors list. First, antenna gain can be 
described through a set of points by using {\it mapping techniques}; and second, the 
use of {\it threading} to explore on each axe direction when using the 
{\it NeighborsGraph} algorithm.

Finally, we consider that our contributions are a first attempt to provide an
asymmetrical communications support within the \Omnet/\INET\,
Framework.




%

\nocite{*}
\bibliographystyle{alpha}
\bibliography{document}
\label{sec:biblio}

\newpage
\tableofcontents

\end{document}